\documentclass[sigconf]{acmart}

\AtBeginDocument{%
  \providecommand\BibTeX{{%
    \normalfont B\kern-0.5em{\scshape i\kern-0.25em b}\kern-0.8em\TeX}}}

\acmPrice{15.00}
\acmISBN{978-1-4503-XXXX-X/18/06}
\usepackage[utf8]{inputenc}

\usepackage{graphicx}
\usepackage{caption}
\usepackage{subcaption}
\usepackage{bbm}
\usepackage{color}
\usepackage{hyperref}
\usepackage{multirow}
\usepackage{enumitem}
\usepackage{algorithm} 
\usepackage{algorithmic}
\usepackage{enumerate}
\usepackage{amsmath}
\usepackage{booktabs}
\usepackage{multirow}
\usepackage{stmaryrd}
\usepackage{mathrsfs} 
\usepackage{color}
\usepackage{balance}

\usepackage[para,online,flushleft]{threeparttable}

\newcommand{\etc}{\textit{etc}.}
\newcommand{\eg}{\textit{e.g.}}
\newcommand{\ie}{\emph{i.e.}}

\usepackage{soul}   
\setlist[itemize]{leftmargin=*}
\usepackage{makecell}

\begin{document}

\title{AlignRec: Aligning and Training in Multimodal Recommendations}
\author{Yifan Liu}
\authornote{Both authors contributed equally to this research.}
\email{sjtulyf123@sjtu.edu.cn}
\orcid{0000-0001-8726-3927}
\affiliation{%
  \institution{Shanghai Jiao Tong University}
  \city{Shanghai}
  \country{China}
}

\author{Kangning Zhang}
\authornotemark[1]
\email{zhangkangning@sjtu.edu.cn}
\orcid{0009-0009-9080-7484}
\affiliation{%
  \institution{Shanghai Jiao Tong University}
  \city{Shanghai}
  \country{China}}

\author{Xiangyuan Ren}
\email{renxiangyuan@xiaohongshu.com}
\orcid{0009-0003-0554-5835}
\affiliation{%
  \institution{Xiaohongshu Inc.}
  \city{Beijing}
  \country{China}}

\author{Yanhua Huang}
\email{yanhuahuang@xiaohongshu.com}
\orcid{0000-0002-3069-4811}
\affiliation{%
  \institution{Xiaohongshu Inc.}
  \city{Shanghai}
  \country{China}}

\author{Jiarui Jin}
\email{jinjiarui@xiaohongshu.com}
\orcid{0000-0001-6458-1586}
\affiliation{%
  \institution{Xiaohongshu Inc.}
  \city{Shanghai}
  \country{China}}

\author{Yingjie	Qin}
\email{yingjieqin@xiaohongshu.com}
\orcid{0009-0005-5078-2201}
\affiliation{%
  \institution{Xiaohongshu Inc.}
  \city{Shanghai}
  \country{China}}

\author{Ruilong	Su}
\email{suruilong@xiaohongshu.com}
\orcid{0009-0003-6695-4536}
\affiliation{%
  \institution{Xiaohongshu Inc.}
  \city{Shanghai}
  \country{China}}

\author{Ruiwen Xu}
\email{rig@xiaohongshu.com}
\orcid{0009-0004-9140-8235}
\affiliation{%
  \institution{Xiaohongshu Inc.}
  \city{Shanghai}
  \country{China}}

\author{Yong Yu}
\email{yyu@apex.sjtu.edu.cn}
\orcid{0000-0003-0281-8271}
\affiliation{%
  \institution{Shanghai Jiao Tong University}
  \city{Shanghai}
  \country{China}
}

\author{Weinan Zhang}
\authornote{Corresponding author.}
\email{wnzhang@sjtu.edu.cn}
\orcid{0000-0002-0127-2425}
\affiliation{%
  \institution{Shanghai Jiao Tong University}
  \city{Shanghai}
  \country{China}
}
\renewcommand{\shortauthors}{Yifan Liu et al.}




\begin{abstract}
With the development of multimedia systems, multimodal recommendations are playing an essential role, as they can leverage rich contexts beyond interactions.
Existing methods mainly regard multimodal information as an auxiliary, using them to help learn ID features; However, there exist semantic gaps among multimodal content features and ID-based features, for which directly using multimodal information as an auxiliary would lead to misalignment in representations of users and items.
In this paper, we first systematically investigate the misalignment issue in multimodal recommendations, and propose a solution named AlignRec.
In AlignRec, the recommendation objective is decomposed into three alignments, namely alignment within contents, alignment between content and categorical ID, and alignment between users and items.
Each alignment is characterized by a specific objective function and is integrated into our multimodal recommendation framework.
To effectively train AlignRec, we propose starting from pre-training the first alignment to obtain unified multimodal features and subsequently training the following two alignments together with these features as input.
As it is essential to analyze whether each multimodal feature helps in training and accelerate the iteration cycle of recommendation models, we design three new classes of metrics to evaluate intermediate performance.
Our extensive experiments on three real-world datasets consistently verify the superiority of AlignRec compared to nine baselines.
We also find that the multimodal features generated by AlignRec are better than currently used ones, which are to be open-sourced in our repository \url{https://github.com/sjtulyf123/AlignRec_CIKM24}.
\end{abstract}



\maketitle


\section{Introduction}\label{sec:intro}
Multimodal recommendations have become one of the essential sources for users to explore content from online services such as social media, especially when ID features are helpless in long-tail items or cold-start scenarios. Several works have demonstrated that utilizing multimodal information (\eg, images and texts) together with categorical features (\eg, ID and category) can gain great success in multiple scenarios~\cite{he2016vbpr,zhang2024drepmrec,wei2020graph,zhou2023bootstrap,zhou2023tale,yu2023multi}.

\begin{figure}[t]
    \centering
    \includegraphics[width=0.8\linewidth]{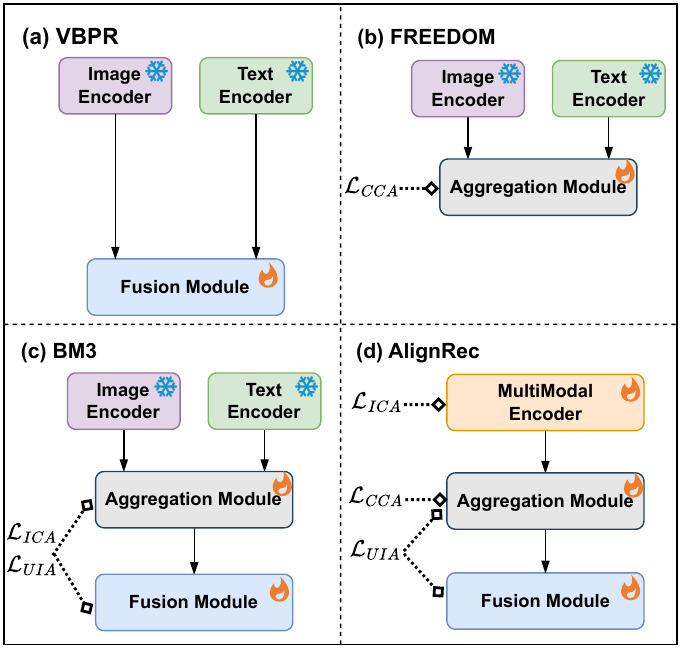}
    \vspace{-3mm}
    \caption{Comparison among VBPR~\cite{he2016vbpr}, FREEDOM~\cite{zhou2023tale}, BM3~\cite{zhou2023bootstrap} and AlignRec. $\mathcal{L}_{ICA}, \mathcal{L}_{CCA}, \mathcal{L}_{UIA}$ are losses for inter-content alignment, content-category alignment and user-item alignment. Dashed lines are scopes of alignment losses.}
    \label{fig:method-diff}
    \vspace{-3mm}
\end{figure}

Meanwhile, alongside the development of recommender systems, the field has witnessed significant advancements in multimodal large language models (mLLMs) like GPT-4V~\cite{achiam2023gpt} and Gemini~\cite{team2023gemini}. One of the key lessons towards generalized mLLMs is the alignment among different content modalities, i.e., viewing different content modalities residing in a unified space~\cite{li2021align,lin2023vila,girdhar2023imagebind,wang2022image}, because the distribution gaps among modalities' representations will significantly harm the cross-modality understanding and make it hard to utilize multimodal features.
However, the alignment issue has not been well-studied in multimodal recommendations. Early works like VBPR~\cite{he2016vbpr} in Fig.~\ref{fig:method-diff}(a) view multimodal information as an auxiliary feature and concatenate it to categorical features~\cite{wei2019mmgcn,wei2020graph}, ignoring the distribution gap between multimodal features and categorical features. 
Some methods~\cite{zhou2023tale,yu2023multi} explore auxiliary losses with multimodal information for deriving better categorical embeddings via graph structures like FREEDOM~\cite{zhou2023tale} in Fig.~\ref{fig:method-diff}(b). But they ignore the inconsistency between image and text. Some self-supervised learning methods~\cite{tao2022self,zhou2023bootstrap} like BM3~\cite{zhou2023bootstrap} in Fig.~\ref{fig:method-diff}(c) train inter-content loss and recommendation-related user-item loss together, leading to an insufficient representation on items' content


In this paper, we propose to leverage ID-based features as a special modality, i.e., \textit{categorical modality}, leading to a unified alignment framework. However, when directly following mLLMs' alignment approaches, there are several remarkable challenges.

\textbf{Challenge 1:} How to perform alignment in multimodal recommendation has not been carefully designed. The input of multimodal recommendation contains not only content modalities like images and text, but also various ID features; And they all remain on their own spaces with different distributions. For content modalities like image and text, they have different semantic information with their own distributions. In experiments, we find that the content and ID feature pair of the same item can also be far away in current methods, which will be discussed in Fig.~\ref{fig:after-align}. Therefore, a unified design for aligning all modalities is required to handle misalignment.

\textbf{Challenge 2:} The popular end-to-end training with joint optimization on alignment and recommendation tasks is hard to learn well. The speeds of learning content modalities and the categorical features are not consistent.
Specifically, the former working on images and texts often requires a large amount of data and time to learn sufficient semantic knowledge~\cite{li2021align}, whereas the latter only needs a few epochs~\cite{zhang2022towards}. Therefore, the speed of learning content modalities is significantly slower than learning categorical modalities. Joint training for multimodal recommendation is hard to deal with this imbalanced rate, introduces high computational overhead, and finally leads to an insufficient representation of items.

\textbf{Challenge 3:} The influence of content modality to multimodal recommendation has not been fully studied. There are metrics designed for mLLMs to determine whether the aligned features are suitable for downstream tasks~\cite{radford2021learning}. However, the multimodal recommendation just incorporates vision and text features into recommendation and evaluates the final performance, which is indirect and does not consider their quality. If the vision or text features are not suitable, more effort is required for re-generation, decreasing the recommendation efficiency. 
We utilize a zero-shot evaluation in Sec.~\ref{sec:inter}, and surprisingly find that the absence of vision features can gain a relatively 12.1\% increase of recall@20.
These observations call for us to directly evaluate the influence of multimodal features to recommendation and select suitable features.

\begin{figure*}[t]
    \centering
    \includegraphics[width=\linewidth]{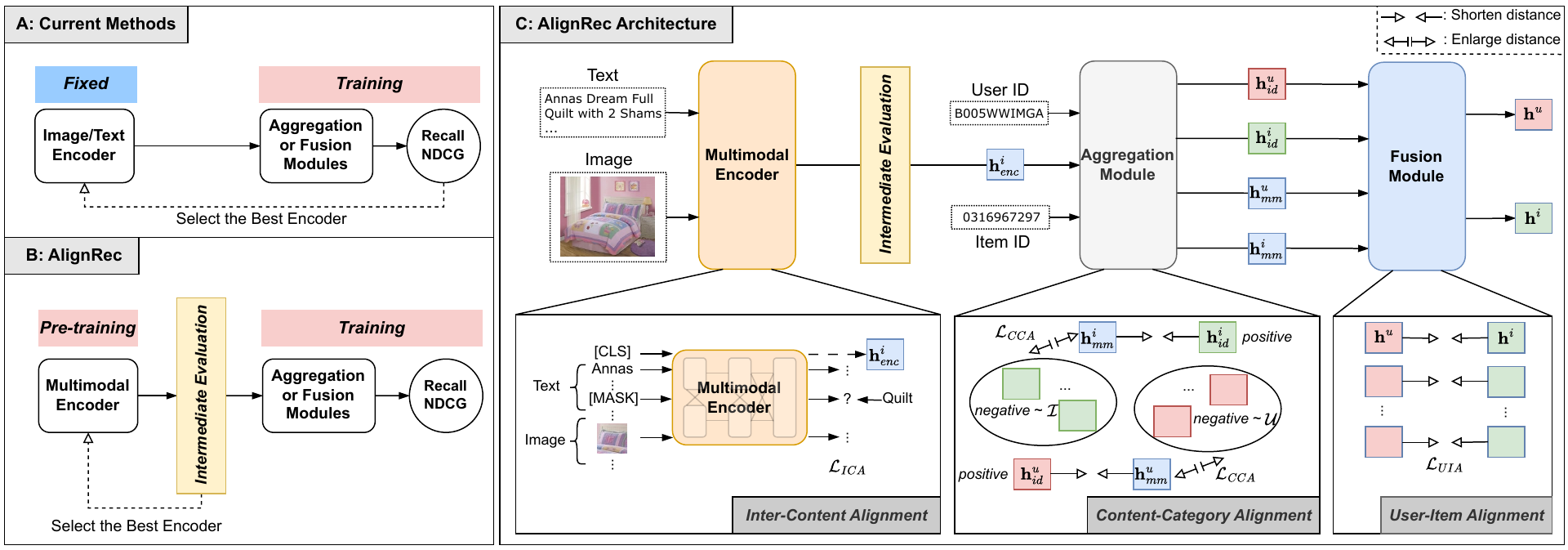}
    \vspace{-5mm}
    \caption{An overview of AlignRec, where user ID, item ID, text and image are input, and user and item representations are output. A and B show the differences between AlignRec and current methods when training, where an intermediate evaluation module and a two-stage training strategy are proposed in AlignRec. C shows the overall architecture of AlignRec.}
    \label{fig:framework}
    \vspace{-4mm}
\end{figure*}

To manage the challenges above, we propose a multimodal recommendation framework \textbf{AlignRec}. 
For \textbf{Challenge 1}, we develop three alignment objectives for different contexts as illustrated in Fig.~\ref{fig:method-diff}(d), respectively. The alignment within contents, namely \textbf{inter-content alignment (ICA)}, aligns different content modality fields (\eg, vision and text) with the help of attention-based cross-modality encoder, and outputs a unified modality representation for one item. The alignment between content and categorical ID, namely \textbf{content-category alignment (CCA)}, utilizes contrastive learning to bridge the gap between multimodal content features and user/item ID-based features for better utilization and adaptation of multimodal information. The \textbf{user-item alignment (UIA)} aligns users and their interacted items via cosine similarity.
For \textbf{Challenge 2}, we propose to train AlignRec by first pre-training the inter-content alignment task, and then training the remaining two alignment tasks together with the recommendation goal. With the pre-training stage utilizing mLLM's capacity to align content modality, the process of jointly learning multi-modality becomes more stable, which leads to a better item/user representation. 
For \textbf{Challenge 3}, we propose three \textbf{intermediate evaluation} protocols, including zero-shot, item-CF and mask modality recommendation, to directly evaluate whether the unified and aligned multimodal features generated via inter-content alignment are helpful for recommendation. It can help us select a better multimodal encoder and reduce the complexity of hyper-parameter search in training. The detailed differences on training and evaluation between AlignRec and current methods will be discussed in Fig.~\ref{fig:framework}.

Our contributions can be concluded as follows:

\begin{itemize}
    \item We systematically analyze that existing multimodal recommendation methods suffer from misalignment issues, and propose a solution AlignRec which can be seamlessly integrated into existing approaches.
    \item We then design to train the AlignRec by first pre-training upon the inter-content alignment task and subsequently further training upon alignment tasks in terms of the recommendation objective, to address the potential learning speed inconsistency between content modality and categorical modality. 
    \item We also provide three classes of new intermediate protocols to evaluate whether the multimodal features are effective, help us select a better encoder, and accelerate the iteration of recommendation models. To the best of our knowledge, we are the first to intermediately evaluate the multimodal recommendation.
\end{itemize}

We conduct experiments on three public datasets and show the state-of-the-art performance of AlignRec compared to nine baselines. 
The intermediate evaluation with in-depth analysis also shows the superior performance of our alignments and training strategies.
Our proceeded multimodal features show better performance than existing ones. 

\section{Related Work}\label{sec:related}
\smallskip
\textbf{Multimodal Recommendation.}
Multimodal recommendation is derived from collaborative filtering (CF) recommendation~\cite{su2009survey}, and incorporates multimodal information for better recommendation. It applies matrix factorization methods~\cite{koren2009matrix}, BPR~\cite{rendle2012bpr} or LightGCN~\cite{he2020lightgcn} and is usually effective in cold-start or long-tail scenarios.

Early methods concatenate vision or text features with ID-based features. VBPR~\cite{he2016vbpr} extends BPR-based recommendation by adding visual features. ACF~\cite{chen2017attentive} applies an attention layer to calculate modalities' contributions to items. MMGCN~\cite{wei2019mmgcn} and GRCN~\cite{wei2020graph} build modality graphs and use GCN~\cite{kipf2016semi} to aggregate information. These methods ignore the misalignment between ID-based and content features, and thus cannot fully exploit multimodal information.

Recently, some methods try to shorten the distance between item's ID and content features. MMGCL~\cite{yi2022multi} applies negative sampling to discover modalities' relationship. SLMRec~\cite{tao2022self} builds self-supervised tasks to enhance the modality representation. BM3~\cite{zhou2023bootstrap} builds text and vision alignment loss by contrastive view. They only focus on the content-category alignment and optimize all losses together, which is hard to converge with insufficient representations.

There are also some methods trying to fuse vision and text information. LATTICE~\cite{zhang2021mining} and FREEDOM~\cite{zhou2023tale} construct modality-similarity graphs using multimodal features and combine these graphs with a pre-defined weight. 
MGCN~\cite{yu2023multi} adopts attention layers to capture the modalities' importance. These methods do not explicitly model the alignment, and will suffer from the modality incompatibility problem (\eg, conflict between text and vision).

\smallskip
\textbf{Vision-Language Pre-training.}
Recently, vision-language pre-training is effective in understanding vision and text information with large-scale data, which is helpful in multimodal recommendation. CLIP~\cite{radford2021learning} proposes to align image and text via contrastive learning and pre-training. ALBEF~\cite{li2021align} designs an image-text contrastive learning task to align image and text features before fusing them. BLIP~\cite{li2022blip} utilizes multi-task learning by building image-text pre-training tasks. BEiT3~\cite{wang2022image} regards image as a foreign language and builds a general make-vision-language task to learn one integrated representation for multimodal features, with the help of multiway transformers~\cite{bao2022vlmo}. 
Due to the complicated structures of these models, many evaluation metrics are proposed. But they are missing in multimodal recommendation. Therefore, most of multimodal recommendation methods still use less powerful features.

\section{Problem Formulation}
Let $u\in \mathcal{U}$ and $i\in \mathcal{I}$ denote the user and item, respectively. The unique ID is denoted as $u_{id}$ and $i_{id}$ . We mainly focus on two modalities, vision and text, and denote item $i$'s raw modality information as $i_v$ and $i_t$, respectively. Besides, we denote users historical behavior data as $\mathbf{R}\in \{0,1\}^{|\mathcal{U}|\times |\mathcal{I}|}$, where each element $r_{u,i}\in \{0,1\}$ represents whether user $u$ clicked item $i$ (1 for clicked and 0 for non-clicked). Based on it, we can construct a user-item bipartite graph $\mathcal{G}=(\mathcal{V},\mathcal{E})$. In $\mathcal{G}$, the vertex set $\mathcal{V}=\mathcal{U} \cup \mathcal{I}$ denotes all items and users; The edge set $\mathcal{E}=\{(u,i)| r_{u,i}=1, u\in\mathcal{U}, i\in \mathcal{I}\}$ represents the interacted users and items.

Based on the notations above, multimodal recommendation's goal is to predict users' preferences on items with a score function $s$ based on the graph $\mathcal{G}$ and multiple features, and then recommend items with high scores to users. In practice, we often utilize Bayesian Personalized Ranking (BPR)~\cite{rendle2012bpr} to optimize this goal~\cite{he2016vbpr, wei2019mmgcn}:
\begin{equation}\label{defi:bpr}
    \max \sum_{(u,i,i')}\ln \sigma\left[ s(f(u),g(i))-s(f(u),g(i'))\right]\,,
\end{equation}
where, $u,i$ is a user-item pair satisfying $r_{u,i}=1$, $i'$ is sampled from $\mathcal{I}$ satisfying $r_{u,i'}=0$. $f:u\rightarrow \mathbb{R}^{d_h}$ and $g:i\rightarrow\mathbb{R}^{d_h}$ are mapping functions to map raw features (ID, images, text, \etc) into dense representations with dimension $d_h$; $s$ is a score function (\eg, inner product); $\sigma$ is the sigmoid function. 

\section{AlignRec}
\subsection{Framework Overview}
The overview of AlignRec and its differences on training and evaluation from current methods are illustrated in Fig.~\ref{fig:framework}. The whole framework consists of three alignments with an intermediate evaluation module. In this section, we will first introduce the architectures, then describe our three alignment objectives, and finally introduce training strategies and intermediate evaluation protocols.

\subsection{Architecture Design}
For generalization, AlignRec only contains three necessary modules: The multimodal encoder module $MMEnc$ aligns the vision and text modality knowledge into one unified representation. The aggregation module $Aggregator$ aggregates the neighbors' information with $\mathcal{G}$. The fusion module $Fuser$ fuses multimodal and ID-based representations for recommendation. 
The goal is to produce item and user representations $\mathbf{h}_{i}$ and $\mathbf{h}_u$ for top-K recommendation.

\subsubsection{Multimodal Encoder Module}
It aims at aligning vision and text of the same item, and outputs a unified multimodal representation $\textbf{h}_{enc}^i$. We propose to adopt a transformer-based multimodal encoder to generate the unified multimodal representation. Formally, suppose $MMEnc$ is a multimodal encoder, and $h_{enc}^i$ is the unified representation for item $i$'s vision and text, then:
\begin{align}\label{eq:mmrec}
    \textbf{h}_{enc}^i = MMEnc(i_v,i_t)\,.
\end{align}

In Eq.~\eqref{eq:mmrec}, the vision-text alignment is optimized through the cross-attention based transformers in $MMEnc$. We utilize the most advanced BEiT3~\cite{wang2022image} as the backbone.
The $MMEnc$ will align vision and text information via cross-modality experts~\cite{bao2022vlmo}. We train it in Sec.~\ref{sec:ica} through the mask-then-predict strategy for a better understanding of both vision and text modalities.

\subsubsection{Aggregation Module}
After obtaining $\mathbf{h}^i_{enc}$, we aggregate items' and users' multimodal and ID-based information with graph $\mathcal{G}$. Generally, suppose $Aggregator$ is an aggregation module. It will output user/item multimodal hidden representations ($\mathbf{h}^u_{mm}$, $\mathbf{h}^i_{mm}$) and ID-based representations ($\mathbf{h}^u_{id}$, $\mathbf{h}^i_{id}$):
\begin{align}\label{eq:aggregation}
    \mathbf{h}^i_{mm}, \mathbf{h}^u_{mm}, \mathbf{h}^i_{id}, \mathbf{h}^u_{id}=Aggregator(\mathbf{h}^i_{enc},i_{id},u_{id}|\mathcal{G})\,.
\end{align}

Specifically, we use LightGCN~\cite{he2020lightgcn,wu2022graph} for aggregation. We adopt an embedding layer to users and items and get $\mathbf{e}^i_{id},\mathbf{e}^u_{id}$. Stacking embeddings of all users and items, we get the matrix $\mathbf{E}_{id}$. Suppose $\hat{\mathbf{A}}$ is the normalized adjacency matrix of $\mathbf{A}$, where $\mathbf{A}=\bigl( \begin{smallmatrix}0 & \mathbf{R}\\ \mathbf{R}^{\intercal} & 0\end{smallmatrix}\bigr)$ is constructed from user-item interaction matrix $\mathbf{R}$. Then, we can apply LightGCN with $L$ layers to aggregate neighbors ID-based information, and obtain $\mathbf{H}_{id}$, which is stacked by $\mathbf{h}^u_{id}$ and $\mathbf{h}^i_{id}$:
\begin{equation}
\begin{aligned}
    \mathbf{H}^{l+1}_{id} &= \hat{\mathbf{A}}\mathbf{H}^l_{id}, \quad \mathbf{H}^0_{id}=\mathbf{E}_{id},\quad l\in \{0,\ldots, L-1\}\,;\\
    \mathbf{H}_{id}&=\frac{1}{L+1}(\mathbf{H}^0+\mathbf{H}^1+\ldots+\mathbf{H}^L)\,.
\end{aligned}
\end{equation}

Secondly, we build the item-side multimodal hidden representation $\mathbf{h}^i_{mm}$ in Eq.~\eqref{eq:item-mm}, where $\odot$ is the element-wise product, $\sigma$ is the sigmoid function, $MLP$ is the multi-layer perceptron. 
After fusing the ID and content knowledge into $\mathbf{h}^i_{con}$, we aggregate neighbors' multimodal information with a similarity matrix $\mathbf{S}$ into $\mathbf{h}^i_{mm}$. $\mathbf{S}$ is built by calculating items' multimodal similarity, and only preserving $K'$ most similarity edges ($K'=10$ following~\cite{yu2023multi}).
\begin{equation}\label{eq:item-mm}
\begin{aligned}
        \mathbf{h}^i_{con}&=\mathbf{e}^i_{id}\odot \sigma \left(MLP(\mathbf{h}^i_{enc})\right)\,,\\
    \mathbf{h}^i_{mm}&=\sum_{j\in \mathcal{I}}\mathbf{S}_{ij}\mathbf{h}^i_{con}\,.
\end{aligned}
\end{equation}

Finally, we build the user-side multimodal hidden representation $\mathbf{h}^u_{mm}$ by aggregating historical behavior of $u$. Denote $\hat{\mathbf{R}}\in \mathbb{R}^{|\mathcal{U}|\times|\mathcal{I}|}$ as the normalized user-item interacted matrix, then:
\begin{align}
    \mathbf{h}^u_{mm}&=\sum_{j\in \mathcal{I}}\hat{\mathbf{R}}_{uj}\mathbf{h}^j_{mm}\,.
\end{align}

\subsubsection{Fusion Module}
The multimodal and ID-based hidden representations contain information of item/user from different fields or aspects. In AlignRec, we use a fusion module $Fuser$ to fuse this information into general item/user representations $\mathbf{h}^i$ and $\mathbf{h}^u$:
\begin{align}\label{eq:fusion}
    \mathbf{h}^i=Fuser(\mathbf{h}^i_{mm}, \mathbf{h}^i_{id}),\quad \mathbf{h}^u=Fuser(\mathbf{h}^u_{mm}, \mathbf{h}^u_{id})\,.
\end{align}

Specifically, we use the element-wise addition operation in Eq.~\eqref{eq:fm-implement}, and finally get a general representation for users and items. This module can also be replaced by more complicated fusion operations like attention in current methods.
\begin{align}\label{eq:fm-implement}
    \mathbf{h}^i=\mathbf{h}^i_{mm}+\mathbf{h}^i_{id},\quad 
    \mathbf{h}^u=\mathbf{h}^u_{mm}+\mathbf{h}^u_{id}\,.
\end{align}

In the top-K recommendation stage, we calculate and rank the score of all user-item pairs given a user $u$: $s(u,i)=\mathbf{h}^i(\mathbf{h}^u)^{\intercal}$. Then, we choose $K$ top-ranked items as recommended items to the user.

\subsection{Three Alignment Objectives}
In this part, we introduce the motivations and designs of our alignments. As discussed before, many studies~\cite{li2021align,lin2023vila} explore the necessity of aligning image and text features and the following gains in cross-modality understanding. Besides, including content modality in alignment can further enhance recommender's ability to deal with long-tail or cold-start scenarios. Regarding ID as a special categorical modality motivates us to analyze the alignment issue in multimodal recommendation and propose the AlignRec.

\subsubsection{Inter-Content Alignment}\label{sec:ica}
In multimodal recommendation scenarios, the item's vision and text modalities usually illustrate the same item from different aspects (\eg, pictures and descriptions of the same product). However, extracting vision and text features from different encoders will result in the distribution gap between content modalities, making it hard for item representations. For example, the vision and text features in commonly used Amazon Sports dataset~\cite{he2016ups} are extracted from VGG and Transformer respectively, which need to be aligned in recommendation.

In Eq.~\eqref{eq:mmrec}, the vision-text alignment is learned through mask-data-modeling techniques: Mask some tokens of one modality (\eg, vision) while predicting the masked token using another modality (text) and remaining tokens in the masked modality (vision). In this way, the $MMEnc$ can attend to both vision and text so that the representation is gradually aligned and fused. The loss is the sum of mask-image-modeling (MIM) and mask-language-modeling (MLM) in Eq.~\eqref{eq:mmrec-loss} introduced in BERT~\cite{devlin2018bert} and BEiT~\cite{wang2022image}. We then obtain the \textbf{CLS} token as the unified multimodal representation $\textbf{h}_{mm}^i$.
\begin{align}\label{eq:mmrec-loss}
    \mathcal{L}_{ICA} = \mathcal{L}_{MIM}+\mathcal{L}_{MLM}\,.
\end{align}

\subsubsection{Content-Category Alignment}
As presented in Fig.~\ref{fig:after-align}(left), one item's multimodal and ID feature without alignment can be far away. It indicates that simply adopting aggregating layers (\eg, MLP or GCN) to acquire multimodal and ID high-level representations separately will cause the distance of content-category pair far away, making it hard for models to distinguish content-category pairs of different items and users. Therefore, we align multimodal and ID hidden representations of the same user or item after aggregation.

According to the $Aggregator$ output in Eq.~\eqref{eq:aggregation}, we utilize the in-batch InfoNCE~\cite{zhao2023augmented,he2020momentum} to model this alignment, guiding our framework learning the difference of positive and negative content-category pairs.
In item-side, the multimodal representation and categorical representation of item $i$ should be close; While the distance between multimodal representation of $i$ and categorical representation of other in-batch items $\tilde{i}$ should be large. And it is the same in user-side.
Denote $\tilde{i}$ and $\tilde{u}$ as in-batch items and users respectively; Denote $\tau$ as the temperature hyper-parameter. Then the content-category alignment task can be optimized through:
\begin{equation}\label{eq:agg-loss}
    \begin{aligned}
    \mathcal{L}_{CCA}&=InfoNCE(i)+InfoNCE(u)\,,\\
    InfoNCE(x) &= -\log \frac{\exp (\mathbf{h}^x_{mm}\cdot (\mathbf{h}^x_{id})^{\intercal}/\tau)}{\sum_{\tilde{x}} \exp (\mathbf{h}^x_{mm}\cdot (\mathbf{h}^{\tilde{x}}_{id})^{\intercal}/\tau)}\,.
\end{aligned}
\end{equation}

Aside from the CCA task, the multimodal representation itself also contains rich information: Similar items' multimodal information should also be similar. However, the multimodal representation $\mathbf{h}^i_{mm}$ will be blindly optimized towards the ID representation $\mathbf{h}^i_{id}$ during the CCA task, leading to the representation collapse~\cite{javaloy2022mitigating}. Therefore, to avoid this problem and keep the similarity relationship, we propose an in-batch regularizer to constrain the similarity between each two multimodal hidden representations $\mathbf{h}^i_{mm}$ stay close to the similarity between $\mathbf{h}^i_{enc}$. Denote $i$ and $j$ as two in-batch items, $C(\mathbf{x},\mathbf{y})$ as the cosine similarity, $\|\cdot\|_2$ as the L2-norm, $sg(\cdot)$ as the stop gradient operator; Then the regularizer is:
\begin{align}\label{eq:sim-loss}
    \mathcal{L}_{REG} = \|C(\mathbf{h}^i_{mm},\mathbf{h}^j_{mm})-sg( C(\mathbf{h}^i_{enc},\mathbf{h}^j_{enc}))\|_{2}\,.
\end{align}

\subsubsection{User-Item Alignment}
After obtaining the general user and item representations in Eq.~\eqref{eq:fusion} via $Fuser$, we try to align the representation space between users and their interacted items. We conduct this user-item alignment because we intend to maximize the prediction of the interacted item $i$ given a user $u$, which will help learn the recommendation goal in Eq.~\eqref{defi:bpr}.
We build the user-item alignment task with a cosine similarity loss between the user final representation $\mathbf{h}^u$ and corresponding interacted item $\mathbf{h}^i$:
\begin{align}\label{eq:ui-loss}
    \mathcal{L}_{UIA}=1-C(\mathbf{h}^i,\mathbf{h}^u)\,.
\end{align}

\subsection{Training and Evaluating AlignRec}
Considering challenge 2 and 3 as mentioned before, we introduce the proposed training strategies, and then show the intermediate evaluation to validate the effectiveness of multimodal features towards recommendation.

\subsubsection{Training Strategies}\label{sec:train}
We use the BPR loss in Eq.~\eqref{defi:bpr} to optimize the recommendation goal, with the inner product as the score function and $i'$ as the negative sampled item, as shown below. It can be generalized to other recommendation goals easily.
\begin{align}\label{eq:bpr-loss}
    \mathcal{L}_{BPR}=\ln \sigma\left[ \mathbf{h}^i(\mathbf{h}^u)^{\intercal}-\mathbf{h}^{i'}(\mathbf{h}^u)^{\intercal}\right]\,.
\end{align}

As discussed in challenge 2, in AlignRec, we have three alignment losses (Eq.~\eqref{eq:mmrec-loss},~\eqref{eq:agg-loss},~\eqref{eq:ui-loss}) together with the recommendation loss (Eq.~\eqref{eq:bpr-loss}) and regularization loss (Eq.~\eqref{eq:sim-loss}) to be optimized. However, optimizing them end-to-end will encounter some problems. 

In view of training speed, the inter-content alignment learns a unified multimodal representation that integrates the information in raw images and text, which requires a large amount of image-text pair data and more training time~\cite{radford2021learning}. On the contrary, the recommendation-related tasks (content-category alignment and user-item alignment) usually converge quickly~\cite{zhang2022towards}. Therefore, training them end-to-end directly will make the recommendation-related tasks learn well before the ICA task, indicating that the recommendation-related tasks cannot obtain the proper multimodal information. 
In view of architecture complexity, the multimodal encoder is composed of complicated transformers, while the recommendation-related tasks consist of simple fully connected or graph layers. As a result, jointly training them is hard for the optimization of the whole network structure.

Considering the difficulties above, in AlignRec, we propose to separate and decouple the whole training process: First, pre-training the inter-content alignment task (named \textit{Pre-training}); Then, training the remaining two alignment tasks together with the recommendation goal (named \textit{Training}) with weight hyper-parameters $\alpha$, $\beta$, and $\lambda$, as written in Eq.~\eqref{eq:train}. The decoupled training processes can enable AlignRec to first generate an aligned and informative multimodal feature for each item with large-scale multimodal data, and then generate suitable user/item representations with the help of content-category alignment and user-item alignment.
\begin{equation}\label{eq:train}
\begin{aligned}
    \textit{Pre-training: }&\min
    \mathcal{L}_{ICA}\,,\\
    \textit{Training: }&\min
    \mathcal{L}_{BPR}+\alpha \mathcal{L}_{CCA}+\beta \mathcal{L}_{UIA}+\lambda \mathcal{L}_{REG}\,.
\end{aligned}
\end{equation}

\subsubsection{Intermediate Evaluation Protocols}\label{sec:inter}
After separating the training process, we evaluate whether our multimodal encoder can generate features helpful for recommendation. The necessity of intermediate evaluation can be summarized as follows:
\begin{itemize}
    \item \textbf{Selecting a better multimodal encoder for recommendation.} We can select a powerful encoder from various encoder models with intermediate evaluation. Then, we can apply it for the downstream recommendation task for better results.
    \item \textbf{Improving multimodal recommendation efficiency.} There are two training strategies in multimodal recommendation: (i) Two-stage training in Eq.~\eqref{eq:train}; (ii) End-to-end training by optimizing losses together. Note that the pre-training stage is more complicated than the second training stage. However, previous methods validate the effectiveness of multimodal features on recommendation metrics.
    If multimodal features are not suitable for recommendation, more effort is required for re-generation.
    \item \textbf{Reducing the complexity of hyper-parameter search in training.} Suppose the pre-training and training stage have $n$ and $m$ hyper-parameter combinations, respectively. Then, the search complexity decreases from $n*m$ (searched together) to $n+m$ (searched separately) with intermediate evaluation.
\end{itemize}
Therefore, we propose three intermediate evaluation protocols in AlignRec to \textbf{directly} evaluate the quality of multimodal features without any ID-based feature or extra fine-tuning. In the following paragraphs, we suppose $\mathbf{h}^i_{mm}$ as item $i$'s multimodal feature.

\paragraph{Zero-Shot Recommendation} It evaluates whether multimodal features can correctly reflect user's interests based on historical interacted items, with three steps:
\begin{itemize}
    \item \textbf{Step I}. Construct user $u$'s interacted items list in time order. Select the most recent interacted item $i$ as the target item, the remaining items as the historical behavior $H_u$.
    \item \textbf{Step II}. Calculate the multimodal feature of user $u$ based on his/her history: $\mathbf{h}^u_{mm}=\frac{1}{|H_u|}\sum_{j\in H_u}\mathbf{h}^j_{mm}$.
    \item \textbf{Step III}. Calculate the similarity between $\mathbf{h}^u_{mm}$ and each item's multimodal feature in the candidate set, and observe whether the target item $i$ can be recalled to $u$ based on top-K similar items.
\end{itemize}
In this protocol, $Recall$ will be larger if the similarity between $\mathbf{h}^{i}_{mm}$ and $\mathbf{h}^u_{mm}$ is large, \ie, historical items' multimodal feature of user $u$ is close to the target item's multimodal feature, with proper reflection on user's interests.

\paragraph{Item-CF Recommendation} It evaluates whether we can correctly recommend items with the CF method via only multimodal features.
\begin{itemize}
    \item \textbf{Step I}. Calculate item-CF score~\cite{10.1145/371920.372071} on each two item pair.
    \item \textbf{Step II}. For each item $j$, select the item with the highest item-CF score as its target item $i$.
    \item \textbf{Step III}. Calculate the similarity between multimodal features of item $j$ and each item in the candidate set, and observe whether $i$ can be recalled on top-K similar items.
\end{itemize}
In this protocol, if the similarity between items' multimodal features is in proportion to the item-CF score, the target item $i$ can be easier to recall, indicating items' multimodal features are well learned.

\paragraph{Mask Modality Recommendation} It measures the difference in performance when we mask items' vision or text modality with a certain percentage. To be specific, we randomly mask $x\%$ of vision of text raw features and generate single-modality representations using $MMEec$. Then, we repeat the steps in the above two protocols and obtain the corresponding recall results.
\begin{table}[t]
    \centering
    \caption{Dataset Statistics.}
    \vspace{-3mm}
\scalebox{0.8}{
\begin{tabular}{c|ccc|ccc}
\toprule
\multirow{2}{*}{Dataset} & \multicolumn{3}{c|}{Raw Data} & \multicolumn{3}{c}{5-core Data} \\ 
\cline{2-7} 
~   &   \#Users    & \#Items      & \#Inters.      &  \#Users     &  \#Items     &   \#Inters.   \\ 
\hline
Baby   &   531,890    &    71,317   &   915,446    &  19,445     &  7,050    & 160,792     \\
Sports   &   1,990,521    &    532,197   &    3,268,695   &  35,598     &  18,357     &  296,337    \\
Electronics     &   4,201,696    &   498,196   & 7,824,482  &   192,403    &  63,001     &   1,689,188      \\ 
\bottomrule
\end{tabular}}
\label{tab:dataset}
\vspace{-5mm}
\end{table}
\section{Experiments}
We conduct experiments to answer following research questions:
\begin{itemize}
    \item \textbf{RQ1}: How does AlignRec perform compared with current multimodal recommendation methods?
    \item \textbf{RQ2}: How does AlignRec perform in intermediate evaluation compared with other pre-extracted multimodal features?
    \item \textbf{RQ3}: How do different components and hyper-parameters of AlignRec affect the multimodal recommendation performance?
    \item \textbf{RQ4}: How does AlignRec perform in long-tail items?
\end{itemize}

\begin{table*}[t]
    \centering
    \caption{Overall performance achieved by different multimodal recommendation methods. * indicates that the improvements are statistically significant compared of the best baseline with $p<0.01$ in t-test.}
    \vspace{-2mm}
\scalebox{0.95}{
\begin{tabular}{c|c|cc|cccccccc}
\toprule
Dataset  & Metric & BPR & LightGCN & VBPR &MMGCN  & GRCN &DualGNN  &BM3  &MGCN  &FREEDOM  & AlignRec \\ \hline
\multirow{4}{*}{Baby} &R@10  &0.0357  &0.0479  &0.0423  &0.0378  &0.0532  &0.0448  &0.0564  &0.0620  &\underline{0.0624}  &\textbf{0.0674*}  \\
    &R@20  &0.0575  &0.0754  &0.0663  &0.0615  &0.0824  &0.0716  &0.0883  &0.0964  &\underline{0.0985}  &\textbf{0.1046*}  \\
    &N@10  &0.0192  &0.0257  &0.0223  &0.0200  &0.0282  &0.0240  &0.0301  &\underline{0.0339}  &0.0324  &\textbf{0.0363*}  \\
    &N@20  &0.0249  &0.0328  &0.0284  &0.0261  &0.0358  &0.0309  &0.0383  &\underline{0.0427}  &0.0416  &\textbf{0.0458*}  \\ 
\hline
\multirow{4}{*}{Sports} &R@10  &0.0432  &0.0569  &0.0558  &0.0370  &0.0559  &0.0568  &0.0656  &\underline{0.0729}  &0.0710  &\textbf{0.0758*}  \\
    &R@20  &0.0653  &0.0864  &0.0856  &0.0605  &0.0877  &0.0859  & 0.0980 &\underline{0.1106}  &0.1077  &\textbf{0.1160*}  \\
    &N@10  &0.0241  &0.0311  &0.0307  &0.0193  &0.0306  &0.0310  &0.0355  &\underline{0.0397}  &0.0382  &\textbf{0.0414*}  \\
    &N@20  &0.0298  &0.0387  &0.0384  &0.0254  &0.0389  &0.0385  &0.0438  &\underline{0.0496}  &0.0476  &\textbf{0.0517*}  \\ 
\hline
\multirow{4}{*}{Electronics} &R@10  &0.0235  &0.0363  &0.0293  &0.0207  &0.0349  &0.0363  &0.0437  &\underline{0.0439}  &0.0382  &\textbf{0.0472*}  \\
    &R@20  &0.0367  &0.0540  &0.0458  &0.0331  &0.0529  &0.0541  &\underline{0.0648}  &0.0643  &0.0588  &\textbf{0.0700*}  \\
    &N@10  &0.0127  &0.0204  &0.0159  &0.0109  &0.0195  &0.0202  &\underline{0.0247}  &0.0245  &0.0209  &\textbf{0.0262*}  \\
    &N@20  &0.0161  &0.0250  &0.0202  &0.0141  &0.0241  &0.0248  &\underline{0.0302}  &0.0298  &0.0262  &\textbf{0.0321*}  \\ 
\bottomrule
\end{tabular}}
\label{tab:overall}
\vspace{-3mm}
\end{table*}

\subsection{Experimental Settings}
\subsubsection{Datasets}
Following~\cite{zhou2023bootstrap,zhou2023tale}, we conduct experiments on three categories of the Amazon review dataset~\cite{he2016ups,mcauley2015image}\footnote{Dataset is available at \url{https://cseweb.ucsd.edu/~jmcauley/datasets/amazon/links.html}.}: (a) \textit{Baby}; (b) \textit{Sports and Outdoors} (denoted as \textit{Sports}); (c) \textit{Electronics}. The Amazon review dataset provides both raw images and text reviews of Amazon products, and each review rating is regarded as a record of positive user-item interaction~\cite{he2016vbpr,zhang2021mining}. The raw data are filtered based on the 5-core setting on both items and users, and previous works mostly use this processed subset for multimodal recommendation. The details of both raw data and filtered 5-core data are in Table~\ref{tab:dataset}, where \#Inters. denotes the user-item interaction number.
As for the multimodal information, previous works use the pre-extracted 4,096-dimensional visual features~\cite{zhou2023bootstrap} and 384-dimensional textual features~\cite{zhou2023mmrec}. However, our framework starts from the raw images and text reviews and produces aligned multimodal features.


\subsubsection{Baselines and Evaluation Metrics}
We compare AlignRec with several state-of-the-art CF-based and multimodal recommendation methods.
For the CF-based methods, we adopt \textbf{BPR}~\cite{rendle2012bpr} and \textbf{LightGCN}~\cite{he2020lightgcn}.
For the multimodal recommendation methods, we adopt \textbf{VBPR}~\cite{he2016vbpr}, \textbf{MMGCN}~\cite{wei2019mmgcn}, \textbf{GRCN}~\cite{wei2020graph}, \textbf{DualGNN}~\cite{wang2021dualgnn}, \textbf{BM3}~\cite{zhou2023bootstrap}, \textbf{MGCN}~\cite{yu2023multi} and \textbf{FREEDOM}~\cite{zhou2023tale}.

For a fair comparison, we follow settings in \cite{zhou2023bootstrap,zhou2023tale,yu2023multi} with the same 8:1:1 splitting strategy on 5-core data for training, validation and testing. We use two widely-used evaluation metrics: $NDCG@K$~\cite{jarvelin2002cumulated} and $Recall@K$~\cite{powers2020evaluation}, denoted as $R@K$ and $N@K$ in tables. In the validation and testing process, we use the all-ranking protocol to calculate metrics based on the scores of users towards all items. We select the best models with the best $Recall@20$ metric on the validation set and then report metrics on the testing set.

\begin{table*}[t]
    \centering
    \caption{Intermediate evaluation of multimodal features through three generation methods.}
    \vspace{-3mm}
    \scalebox{0.9}{
\begin{tabular}{c|c|ccc|ccc|ccc|ccc}
\toprule
\multirow{2}{*}{Dataset} & \multirow{2}{*}{Metric} & \multicolumn{3}{c|}{Zero-Shot} & \multicolumn{3}{c|}{Mask 50\% Vision} & \multicolumn{3}{c|}{Mask 50\% Text} & \multicolumn{3}{c}{Item-CF} \\ \cline{3-14} 
 &  &Amazon  & CLIP  &AlignRec  &Amazon  &\ CLIP  &AlignRec  &Amazon  & CLIP  &AlignRec  &Amazon  & CLIP &AlignRec  \\ \hline
 \multirow{3}{*}{Baby}&R@10  &0.0031  &0.0093  &\textbf{0.0113}  &0.0039  &0.0084  &\textbf{0.0089}  &0.0019  &0.0058  &\textbf{0.0067}  &0.0237  &0.0541  &\textbf{0.0782}  \\
     &R@20  &0.0058  &0.0158  &\textbf{0.0231}  &0.0065  &0.0146  &\textbf{0.0174}  &0.0044  &0.0117  &\textbf{0.0131}  &0.0318  &0.0714  &\textbf{0.1005}  \\
     &R@50  &0.0146  &0.0323  &\textbf{0.0470}  &0.0187  &0.0292  &\textbf{0.0365}  &0.0145  &0.0236  &\textbf{0.0269}  &0.0440  &0.0987  &\textbf{0.1350}  \\ \hline
 \multirow{3}{*}{Sports}&R@10  &0.0056  &0.0136  &\textbf{0.0165}  &0.0041  &\textbf{0.0124}  &\underline{0.0118}  &0.0036  &0.0044  &\textbf{0.0052}  &0.0268  &0.0565  &\textbf{0.0772}  \\
       &R@20  &0.0081  &0.0226  &\textbf{0.0268}  &0.0073  &0.0206  &\textbf{0.0216}  &0.0056  &0.0074  &\textbf{0.0085}  &0.0335  &0.0756  &\textbf{0.1059}  \\
       &R@50  &0.0120  &0.0418  &\textbf{0.0457}  &0.0111  &0.0384  &\textbf{0.0401}  &0.0091  &\textbf{0.0198}  &\textbf{0.0198}  &0.0454  &0.1028  &\textbf{0.1497}  \\ 
 \bottomrule
\end{tabular}
    }
    \label{tab:intermediate}
    \vspace{-2mm}
\end{table*}

\begin{table*}[t]
    \centering
    \caption{Replacement of Amazon features with new features generated by our framework.}
    \vspace{-3mm}
\scalebox{0.9}{
\begin{tabular}{c|c|ccc|ccc|ccc}
\toprule
\multirow{2}{*}{Dataset} & \multirow{2}{*}{Metric} & \multicolumn{3}{c|}{VBPR} & \multicolumn{3}{c|}{BM3} & \multicolumn{3}{c}{FREEDOM} \\ \cline{3-11} 
 &  & Amazon & CLIP & AlignRec & Amazon & CLIP & AlignRec & Amazon & CLIP & AlignRec \\ \hline
\multirow{5}{*}{Baby} & R@10 &0.0423  &0.0467  &\textbf{0.0495}  &0.0532  &0.0538  &\textbf{0.0545}  &0.0624  &0.0629  &\textbf{0.0653}  \\
 & R@20 &0.0663  &0.0737  &\textbf{0.0768}  &0.0850  &0.0851  &\textbf{0.0864}  &0.0985  &0.0987  &\textbf{0.0999}  \\
 & N@10 &0.0223  &0.0253  &\textbf{0.0265}  &0.0277  &0.0282  &\textbf{0.0289} &0.0324  &0.0337  &\textbf{0.0349}  \\
 & N@20 &0.0284  &0.0322  &\textbf{0.0337}  &0.0359  &0.0366  &\textbf{0.0369}  &0.0416  &0.0428  &\textbf{0.0438}  \\ 
  & MEM $\downarrow\%$ &-  &13.90  &\textbf{16.65}  &-  &7.28  &\textbf{10.45}  &-  &24.60  &\textbf{27.54}  \\ 
 \hline
\multirow{5}{*}{Sports} & R@10 &0.0558  &0.0554  &\textbf{0.0570}  &0.0622  &0.0629  &\textbf{0.0655}  &0.0710  &0.0716  &\textbf{0.0724}  \\
 & R@20 &0.0856  &0.0865  &\textbf{0.0885}  &0.0962  &0.0965  &\textbf{0.0983}  &0.1077  &0.1103  &\textbf{0.1110}  \\
 & N@10 &0.0307  &0.0297  &\textbf{0.0310}  &0.0336  &0.0347  &\textbf{0.0362}  &0.0382  &0.0383  &\textbf{0.0394}  \\
 & N@20 &0.0384  &0.0381  &\textbf{0.0392}  &0.0424  &0.0433  &\textbf{0.0447}  &0.0476  &0.0483  &\textbf{0.0490}  \\ 
   & MEM $\downarrow\%$ &-  &22.28  &\textbf{26.23}  &-  &13.02  &\textbf{16.59}  &-  &22.17  &\textbf{26.15}  \\ 
 \bottomrule
\end{tabular}}
    \label{tab:replacement}
\vspace{-3mm}
\end{table*}

\subsubsection{Implementation Details}
We implement our framework with PyTorch based on MMRec~\cite{zhou2023mmrec}, a unified publicly repository specifically designed for implementing new multimodal recommendation methods. In the inter-content alignment task, we train our multimodal encoder on the large-scale \textbf{raw data} to obtain a unified knowledge between image and text, with the help of BEiT3 backbone in TorchScale~\cite{torchscale}\footnote{Variants of BEiT3 are at \url{https://github.com/microsoft/unilm/tree/master/beit3}.}. We train it on the raw data rather than the filtered 5-core data because the latter do not have enough image-text pairs for alignment. After that, we use the generated multimodal features to train the remaining alignment objectives together with the BPR loss and evaluate the results on \textbf{5-core data}, to make our framework be consistent with current baselines.

\subsection{Overall Performance (RQ1)}
Table~\ref{tab:overall} shows the overall top-10 and top-20 performance of our proposed AlignRec framework and other baselines on three datasets. Following the results, observations are concluded as follows:
\begin{itemize}
    \item AlignRec outperforms current multimodal recommendation baselines among all three datasets. Specifically, the improvement of our framework compared with the second best method is $6.19\%, 4.88\%, 8.02\%$ in $Recall@20$. Note that our framework focuses on the alignment problem and just uses some simple neural network structures. Therefore, the improvement is mainly contributed by our three alignments and the training strategies, which demonstrate the effectiveness of our framework. Moreover, the improvement in the ranking-related $NDCG$ metric also shows that AlignRec can correctly rank items by users' interests.
    \item Some baselines like FREEDOM perform well in \textit{Baby} and \textit{Sports} dataset but fail to achieve the similar performance in \textit{Electronics}. On the contrary, BM3 achieves the second best performance in \textit{Electronics} dataset. It is because the \textit{Electronics} dataset contains a large amount of users and items, and requires a better utilization usage of multimodal features to help ID-based features learn item-user relationships. However, FREEDOM only regards multimodal information as an indicator to build the item-item graph, without carefully incorporating it into the top-K recommendation process. AlignRec adopts three alignments to ensure that the multimodal information is suitable for recommendation and can properly reflect item's properties, and thus achieves better performance.
    \item In general, graph-based methods (\eg, LightGCN, MMGCN) is better than those that non-graph methods (\eg, VBPR). It is reasonable since the user-item interaction graph indicates users' interests and similarities, which is helpful for the generalization of recommendation methods. 
    In addition, we also find that multimodal methods generally outperform than traditional methods with ID features only. This observation demonstrates the necessity of adding modality knowledge to recommendation.
\end{itemize}

\subsection{Intermediate Evaluation (RQ2)}\label{sec:eval}
We also conduct the experiments introduced in intermediate evaluation protocols to validate the effectiveness of our $MMEnc$ with the inter-content alignment task, and determine whether the generated multimodal representations are suitable for recommendation. In this part, we compare multimodal representations generated by our framework with other pre-extracted features on \textit{Baby} and \textit{Sports} datasets. We denote \textbf{Amazon} as the original vision and text features collected from datasets, which are used by previous methods~\cite{he2016vbpr,wei2019mmgcn,yu2023multi}; Denote \textbf{CLIP} as the vision and text features extracted from the finetuned CLIP model. Since our framework only generates one unified multimodal feature for one item, we fuse vision and text features via concatenation on \textbf{Amazon} and \textbf{CLIP} features for fair comparison. We present the $Recall$ metric of zero-shot, item-CF and mask modality recommendation in Table~\ref{tab:intermediate}. The mask modality recommendation is conducted by masking \textbf{50\%} items' modality features based on zero-shot settings.

From the results in Table~\ref{tab:intermediate}, we can find that: 
(i) Compared with original Amazon multimodal features, which are generated by CNN and Transformer separately, features generated by CLIP and AlignRec enjoy better zero-shot and item-CF performance in the $Recall$ metric. It is because Amazon features do not consider the semantic gap between vision and text information; while CLIP uses contrastive learning and AlignRec utilizes mask-then predict strategy with cross-attention to do the inter-content alignment, and thus have a better representation of items. The performance gap between AlignRec and CLIP also demonstrates the effectiveness of our $MMEnc$. 
(ii) When we mask some items' vision or text features, AlignRec also outperforms Amazon and CLIP. This mask modality recommendation evaluation simulates the real recommendation scenario where the modality raw information is missing. Therefore, our framework is more robust in the missing modality scenario. It is attributed to the inter-content alignment to learn multimodal knowledge from both vision and text information.
(iii) Compared with masking vision and masking text, we find that masking text will bring about a greater decrease in $Recall$ in general. It indicates that text modality is more important than vision, and paying more attention to modeling text information may be helpful for multimodal recommendation.
(iv) We surprisingly find that when we mask $50\%$ items' vision modality in Amazon features, the performance increases (\eg, 0.0058 to 0.0065 in $Recall@20$). It is because the vision feature collected by Amazon cannot correctly represent items, which has been partly discussed in~\cite{zhou2023tale}. 

To further investigate the reasons for this interesting phenomenon (iv), we try to only use Amazon text features and ID features to train several baselines (\ie, do not use vision features), and compare the performance with their original results. Table ~\ref{tab:amazon-vision} shows the detailed results on Baby dataset. We can observe that when we train baselines without Amazon vision features, the performance is still close to or even better than the original results. This is because the Amazon vision features contain less valuable information and do not contribute much to the final recommendation.

\begin{table}[t]
    \centering
    \caption{Analysis of training baselines without vision features on the Amazon Baby dataset.}
    \vspace{-2mm}
    \scalebox{0.8}{
    \begin{tabular}{c|cccc}
    \toprule
         Baseline& R@10&R@20&N@10&N@20 \\
         \midrule
         VBPR& 0.0423&0.0663&0.0223&0.0284\\
         VBPR w/o Vision&0.0458&0.0723&0.0242&0.0310 \\
                  \midrule
         BM3& 0.0564&0.0883&0.0301&0.0383\\
         BM3 w/o Vision&0.0565&0.0880&0.0299&0.0385 \\
                  \midrule
         FREEDOM&0.0624&0.0985&0.0324&0.0416 \\
         FREEDOM w/o Vision&0.0620&0.0993&0.0326&0.0422 \\
         \bottomrule
    \end{tabular}}
    \label{tab:amazon-vision}
    \vspace{-2mm}
\end{table}

\subsection{In-depth Analysis (RQ3)}
\subsubsection{Training Baselines with Our Features}
The intermediate evaluation experiments already show that the inter-content alignment of AlignRec can generate more suitable multimodal features for recommendation. Furthermore, we wonder whether the multimodal features generated by our framework can be applied to existing methods with improvement. Following this idea, we replace the currently used Amazon features with CLIP features and our proposed new features, and train three typical multimodal recommendation methods: VBPR, BM3, and FREEDOM. Please note that: (1) Amazon and CLIP contain both vision and text features, but our framework only contains one unified feature for one item. Therefore, we simplify some structures designed for two modalities in these methods (\eg, merging vision and text MLPs into one MLP) when training AlignRec features. (2) To test how different features actually perform, we freeze the multimodal features in BM3 since this method heavily relies on them.

From the performance in Table~\ref{tab:replacement}, we find that our features can also achieve better performance than Amazon and CLIP multimodal features on three baselines, demonstrating the effectiveness of our inter-content alignment task and its role for recommendation. Besides, we calculate the memory decrease with regard to using Amazon features, denoted as MEM $\downarrow\%$ in Table~\ref{tab:replacement}. In terms of memory efficiency, our features also achieve the best performance. This is because we only generate one 768-dim multimodal feature for one item, and do not need structures specifically designed for fusing vision and text information. Following these observations, we conclude that features generated by AlignRec is \textbf{small but efficient}, which can be used for future multimodal recommendations.

\begin{table}[t]
    \centering
    \caption{Ablation study on different variants.}
    \vspace{-3mm}
    \scalebox{0.85}{
\begin{tabular}{c|c|cccc}
\toprule
Dataset & Variant & R@10 & R@20 & N@10 & N@20 \\ \hline
\multirow{6}{*}{Baby} &AlignRec  &\textbf{0.0674}  &\textbf{0.1046}  &\textbf{0.0363}  &\textbf{0.0458}  \\
 &w/o CCA  &0.0627  &0.0972  &0.0339  &0.0428  \\
 &w/o UIA  &0.0640  &0.0998  &0.0348  &0.0439  \\
 &w/o REG  &0.0604  &0.0946  &0.0323  &0.0411  \\
 &w/o Vision  &0.0611  &0.0962  &0.0333  &0.0423  \\
 &w/o Text  &0.0580  &0.0925  &0.0311  &0.0400  \\ \hline
\multirow{6}{*}{Sports} &AlignRec  &\textbf{0.0758}  &\textbf{0.1160}  &\textbf{0.0414}  &\textbf{0.0517 } \\
 &w/o CCA  &0.0663  &0.1028  &0.0357  &0.0451  \\
 &w/o UIA  &0.0726  &0.1110  &0.0392  &0.0490  \\
 &w/o REG  &0.0690  &0.1059  &0.0368  &0.0463  \\
 &w/o Vision  &0.0747  &0.1124  & 0.0401 &0.0499  \\
 &w/o Text  &0.0677  &0.1022  &0.0372  &0.0461  \\ \bottomrule
\end{tabular}}
\label{tab:ablation}
\vspace{-5mm}
\end{table}

\subsubsection{Ablation Study}
We ablate AlignRec by removing important components in it and observe performance on \textit{Baby} and \textit{Sports} datasets. \textbf{w/o CCA} and \textbf{w/o UIA} denote our framework without content-category alignment and without user-item alignment, respectively. \textbf{w/o REG} denotes removing the modality similarity regularizer in Eq.~\eqref{eq:sim-loss}. \textbf{w/o Vision} and \textbf{w/o Text} denote removing the vision or text modality from input.

From Table~\ref{tab:ablation}, we can conclude that all components make contributions to AlignRec. The performance gap between AlignRec and w/o CCA or w/o UIA demonstrates the effectiveness of our proposed alignment tasks. By aligning multimodal information, user ID, and item ID, our framework is capable of properly incorporating helpful multimodal representations into recommendation and alleviating distribution inconsistency. Besides, removing the similarity regularizer also decreases the performance, which shows that keeping the multimodal similarity relationships among items is necessary for preserving semantic information when optimization. The results on w/o Vision and w/o Text variants indicate that both item's image and text description are useful for building a comprehensive understanding of the item and helping recommendation. 

\subsubsection{Hyper-parameter Study}
\begin{figure}[t]
    \centering
    \includegraphics[width=\linewidth]{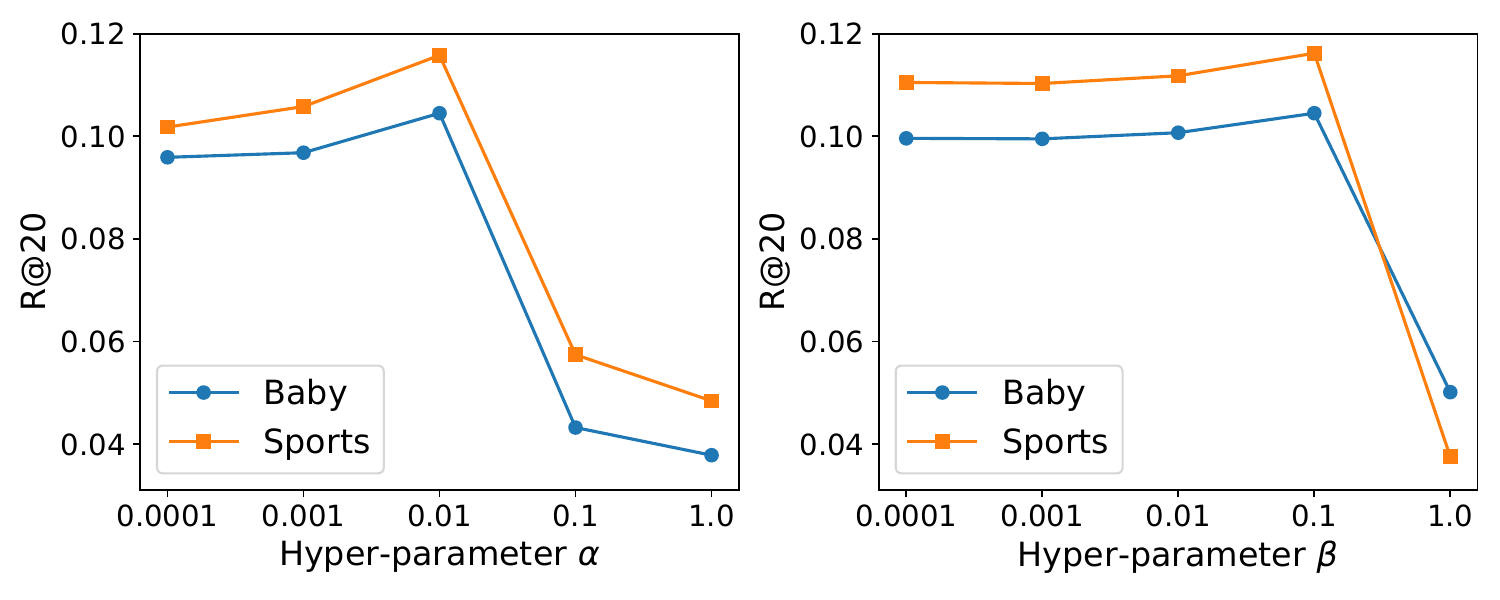}
    \vspace{-7mm}
    \caption{Hyper-parameter study on two alignment weights.}
    \vspace{-3mm}
    \label{fig:hyper-parameter}
\end{figure}

We conduct a study on two hyper-parameters in AlignRec to show their sensitivity: content-category alignment weight $\alpha$ and user-item alignment weight $\beta$, as defined in Eq~\eqref{eq:train}. The results are illustrated in Fig.~\ref{fig:hyper-parameter}. We find that, when the weight $\alpha$ is increasing from 0.0001 to 0.01, the $Recall@20$ also gradually increases to its peak, indicating that the content-category alignment task is working well by the increased importance. However, if the weight is too large (\eg, 0.1), AlignRec will mostly focus on this alignment task but ignore other goals, and the performance will drop dramatically. The same phenomenon is also found in user-item alignment task weight $\alpha$ with peak performance in $\beta=0.1$. 

\subsubsection{Representation Visualization on Alignment}
\begin{figure}
    \centering
    \includegraphics[width=0.49\linewidth]{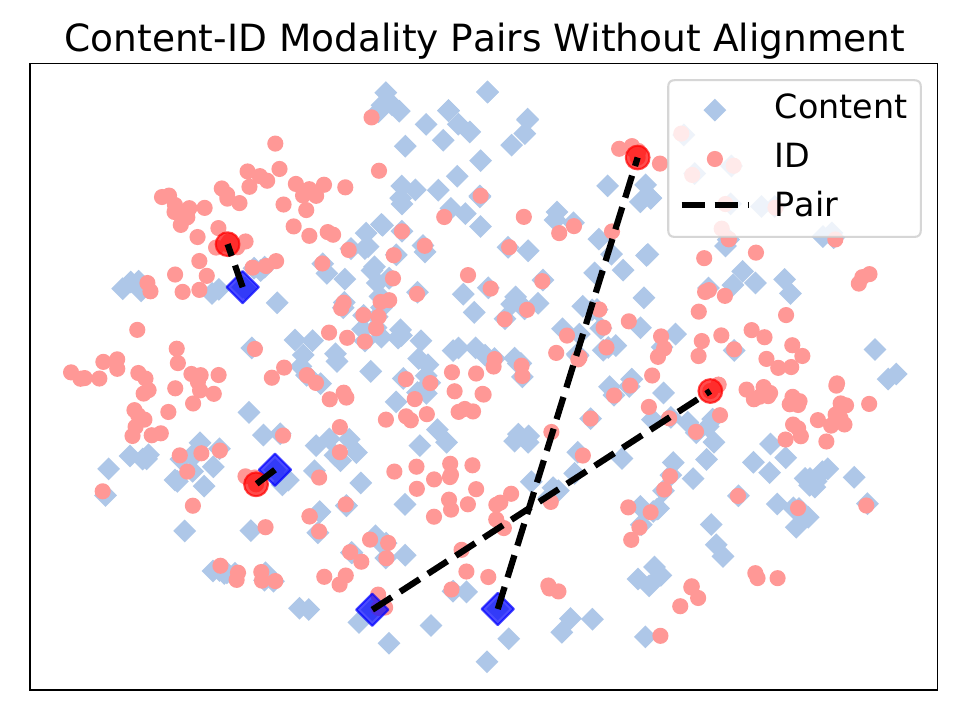}
    \includegraphics[width=0.49\linewidth]{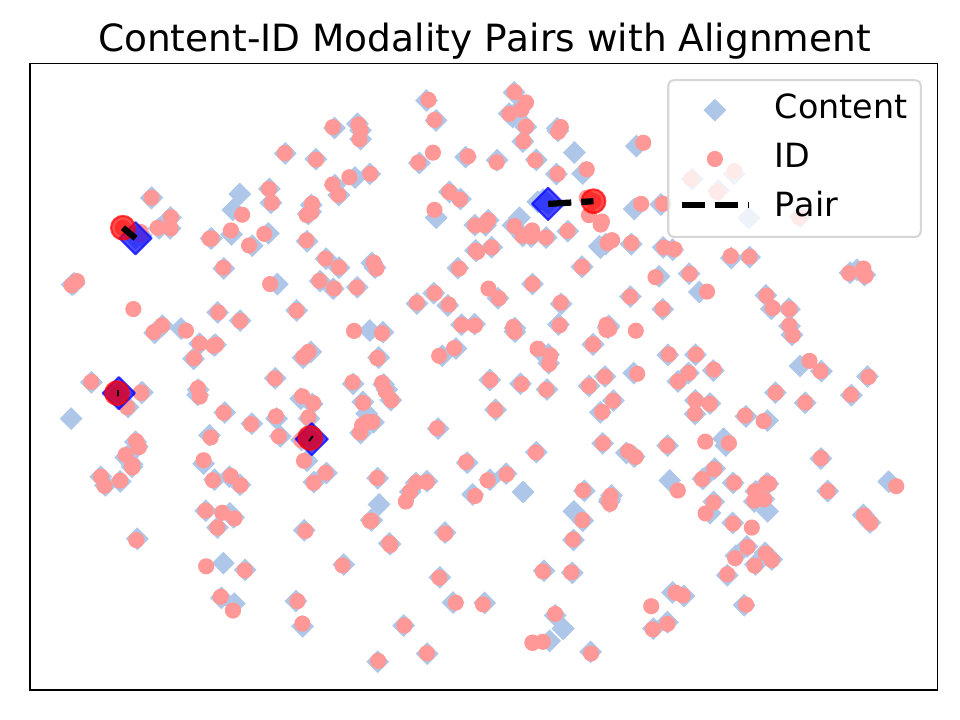}
    \vspace{-4mm}
    \caption{The t-SNE results of content and ID modality feature pairs with and without alignment.}
    \label{fig:after-align}
    \vspace{-3mm}
\end{figure}

To further investigate in the impact of our alignment strategies, we visualize items' content and ID-based representations (\ie, $\mathbf{h}^i_{mm}$ and $\mathbf{h}^i_id$ in Eq.~\eqref{eq:aggregation}) of well-trained AlignRec with the t-SNE method. The results on Amazon Sports dataset are presented in Fig.~\ref{fig:after-align}, where the left figure shows the results on a well-trained VBPR model without alignment, and the right figure shows the results on AlignRec with alignment. 
We can observe that, when there are no alignment losses, the content and ID feature pair of the same item can be far away. However, the content-ID pairs become similar after alignment. We also calculate the cosine similarity of ID and content representations before and after alignment. The similarity before alignment is \textbf{0.5035}, while the similarity after alignment is \textbf{0.8631}. They validate the necessity and effectiveness of our alignment strategies.

\begin{figure}[t]
    \centering
    \includegraphics[width=\linewidth]{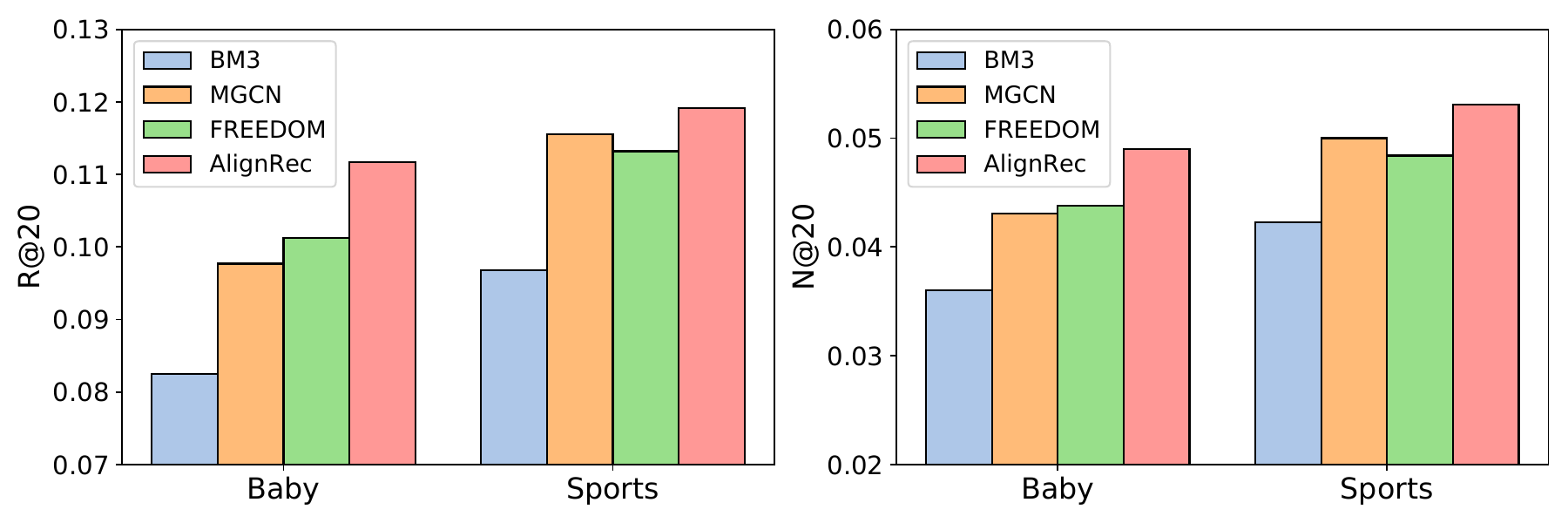}
    \vspace{-8mm}
    \caption{Results of long-tail items recommendation.}
    \vspace{-5mm}
    \label{fig:long-tail}
\end{figure}

\subsection{Long-Tail Items Recommendation (RQ4)}
As discussed before, one important application of multimodal recommendation is to recommend items that have only been browsed by few users to people who may be interested in them, \ie, long-tail items. In this scenario, the items' ID-based features can not provide much information since they do not have enough interactions. Therefore, multimodal information like items' images and descriptions serve as a critical role in discovering potential audiences because of rich semantic knowledge and generalizability. In this paper, we also conduct a study on long-tail items. Specifically, we only focus on items in testing set that have been interacted with less than four times and regard them as long-tail items. We record their performance using different multimodal recommendation methods. 

From the results in Fig.~\ref{fig:long-tail}, we can discover that, AlignRec outperforms current methods (BM3, MGCN and FREEDOM). It indicates that our framework can obtain useful multimodal information from hot items and generalize it to the recommendation process of long-tail items, with the help of three alignments. We also notice that BM3 performs worse than MGCN and FREEDOM. It is reasonable since MGCN and FREEDOM both utilize the modality-based similarity graph to capture multimodal relationships among items, and thus they can have better generalizability to long-tail items.

\section{Conclusion}
In this paper, we focus on the multimodal recommendation in the view of alignment, and propose AlignRec. In AlignRec, we propose to utilize three alignment tasks to help the recommendation goal: The inter-content alignment tries to align and fuse the information between vision and text modality; The content-category alignment manages to eliminate the gap between the multimodal and ID-based representations; The user-item alignment aims to make interacted user and item close to each other. We also propose an efficient and practical training strategy for our framework, followed by new evaluation protocols to validate the effectiveness of multimodal representations. Comprehensive experiments demonstrate the superior performance of AlignRec. In the future, it would be interesting to evaluate
and deploy AlignRec on an online social media platform.

\section*{Acknowledgement}
The Shanghai Jiao Tong University team is partially supported by National Natural Science Foundation of China (62177033, 62076161) and Shanghai Municipal Science and Technology Major Project (2021SHZDZX0102).

\bibliographystyle{ACM-Reference-Format}
\balance
\bibliography{myref}

\newpage
\section*{Appendix}
\subsection*{Details of Mask-Language-Modeling and Mask-Image-Modeling}

We present the detailed loss formulation of mask-language-modeling (MLM) and mask-image-modeling (MIM), which are derived from the definitions in BERT~\cite{devlin2018bert}.

Suppose an item has text $t$ and image $v$. For $t$, we tokenize it and get $N$ text tokens $\mathcal{T}=\{t_i\}_{i=1}^N$ ($N$ is the pre-defined max length). Similarly, we split $v$ into $M$ image patches and tokenize it to $M$ visual tokens $\mathcal{V}=\{v_j\}_{j=1}^M$. We randomly mask approximately 15\% text tokens and 40\% image tokens, where the masked positions are denoted as $\mathcal{P}^t \in \{1,...,N\}^{0.15N}$ and $\mathcal{P}^v \in \{1,...,M\}^{0.4M}$, respectively. Next we replace the masked tokens with a unique text token $e^t$ and image token $e^v$, with learnable embeddings. Then the corrupted text tokens are $\hat{\mathcal{T}}=\{t_i| i \notin\mathcal{P}^t \}_{i=1}^N\cup\{e^t| i \in\mathcal{P}^t \}_{i=1}^N$; The corrupted image tokens are $\hat{\mathcal{V}}=\{v_j| j \notin\mathcal{P}^v \}_{j=1}^M\cup\{e^v| j \in\mathcal{P}^v \}_{j=1}^M$.
The goal of inter-content alignment is to: (i) Predict the correct text tokens $z_i^t$ given corrupted text tokens $\hat{\mathcal{T}}$ and all image tokens $\mathcal{V}$, by minimizing $\mathcal{L}_{MLM}$; (ii) Predict the correct image tokens $z_j^v$ given corrupted image tokens $\hat{\mathcal{V}}$ and all text tokens $\mathcal{T}$, \textit{i.e.}, by minimizing $\mathcal{L}_{MIM}$. Formally, we use the maximum likelihood estimation to formulate them, where $p$ is an encoder with a softmax classifier:
\begin{align*}
\mathcal{L}_{MLM}&= \max \mathbb{E}_{\mathcal{P}^t}\left[ \sum_{i\in \mathcal{P}^t}\log p(z_i^t|\hat{\mathcal{T}}\cup\mathcal{V})\right]\,,\\
\mathcal{L}_{MIM}&= \max \mathbb{E}_{\mathcal{P}^v}\left[ \sum_{j\in \mathcal{P}^v}\log p(z_j^v|\hat{\mathcal{V}}\cup\mathcal{T})\right]\,.
\end{align*}

Intuitively, $\mathcal{L}_{MLM}$ and $\mathcal{L}_{MIM}$ together aim to capture the relationship between image and text modality of an item. Then our MM encoder fuses the image and text representations to obtain the representation of the item.

\subsection*{Compared Baselines}
\begin{itemize}
    \item \textbf{BPR}~\cite{rendle2012bpr}: It leverages the historical interactions between users and items to capture latent factors that represent users and items. It predicts user preferences by measuring the similarity between these representations. This model introduces the widely used BPR loss function.
    \item \textbf{LightGCN}~\cite{he2020lightgcn}:
This is the most popular GCN-based collaborative filtering method, which simplifies the design of GCN to make it more appropriate for the recommendation. 
    \item \textbf{VBPR}~\cite{he2016vbpr}:
This model integrates the visual features and ID embeddings of each item as its representation. To be fair, we concatenate both vision and text features as the multimodal feature when we learn the representations. The BPR loss is used here to learn the user preference. 
    \item \textbf{MMGCN}~\cite{wei2019mmgcn}:
The model learns the modality-specific preference by constructing the modal-specific graphs for each modality and utilizing those user-item graphs to learn the representations in each modality. Finally, the model aggregates all the modality-specific representations to get the final user and item embeddings. 
    \item \textbf{GRCN}~\cite{wei2020graph}:
This model uses the same user-item graphs as previous GCN-based models, but it identifies false-positive edges in the graph and removes them. The refined graph is then used to generate new representations by performing aggregation and information propagation.
    \item \textbf{DualGNN}~\cite{wang2021dualgnn}:
This model introduces a novel user-user co-occurrence graph utilizing representations learned from modality specific graphs and incorporates neighbor representations.
    \item 
\textbf{BM3}~\cite{zhou2023bootstrap}: This method simplifies the self-supervised approach. It removes the requirement of randomly sampled negative examples. It employs the dropout technique to generate contrastive views and utilizes three contrastive loss functions to optimize the resulting representations.
\item \textbf{MGCN}~\cite{yu2023multi}:
This model addresses modality noise by purifying modality features with item behavior information. This model aggregates multi-view representations for recommendation.
\item \textbf{FREEDOM}~\cite{zhou2023tale}: This model mines the latent structure between items by learning an item-item graph and freezing the graphs before training. It introduces degree-sensitive edge pruning techniques to remove noise from the user-item interaction graph.
\end{itemize}

\begin{table}[t]
    \centering
    \caption{Detailed hyper-parameters when training AlignRec.}
    \begin{tabular}{l|c}
    \toprule
       Hyper-parameters  & Values \\
    \midrule
    Embedding dimension $d_e$&64\\
    Multimodal feature dimension & 768\\
    MLP hidden size & 64 \\
    Aggregator hidden size& 64\\
    Fuser hidden size & 64 \\
    GCN layers on interaction graph $\mathcal{G}$ & 2\\
    GCN layers on modality similarity graph $\mathcal{S}$ & 1 \\
    kNN $K'$ when filtering $\mathcal{S}$ & 10 \\
    \midrule
       Learning rate  & \{3e-4,1e-4\} \\
       Learning rate scheduler & LambdaLR \\
       Batch size & 2048 \\
       Optimizer & Adam \\
       CCA weight $\alpha$ &0.01 \\
       UIA weight $\beta$ & 0.1 \\
       Similarity regularizer weight $\lambda$ &\{0.1,0.2,0.3\} \\
       InfoNCE temperature $\tau$ & 0.2\\
    \bottomrule
    \end{tabular}
    \label{tab:appendix-hyper}
\end{table}

\begin{table*}[t]
    \centering
    \caption{Ablation study with mean and std values.}
    \vspace{-2mm}
\begin{tabular}{c|c|cccc}
\toprule
Dataset & Variant & R@10 & R@20 & N@10 & N@20 \\ \hline
\multirow{6}{*}{Baby} &AlignRec  &\textbf{0.0673$\pm$ 0.0009}  &\textbf{0.1039$\pm$0.0011}  &\textbf{0.0365$\pm$0.0003}  &\textbf{0.0458$\pm$0.0003}  \\
 &w/o CCA  &0.0622$\pm$0.0008  &0.0977$\pm$0.0011  &0.0339$\pm$0.0005  &0.0430$\pm$0.0006  \\
 &w/o UIA  &0.0639$\pm$0.0007  &0.0996$\pm$0.0010  &0.0345$\pm$0.0004  &0.0440$\pm$0.0005  \\
 &w/o REG  &0.0603$\pm$0.0010  &0.0940$\pm$0.0012  &0.0321$\pm$0.0006  &0.0412$\pm$0.0005  \\
 &w/o Vision  &0.0609$\pm$0.0009  &0.0960$\pm$0.0013  &0.0333$\pm$0.0003 &0.0422$\pm$0.0006  \\
 &w/o Text  &0.0577$\pm$0.0005  &0.0924$\pm$0.0010  &0.0310$\pm$0.0003  &0.0399$\pm$0.0005  \\ \hline
\multirow{6}{*}{Sports} &AlignRec  &\textbf{0.0764$\pm$0.0006}  &\textbf{0.1158$\pm$0.0007}  &\textbf{0.0416$\pm$0.0003  }&\textbf{0.0517 $\pm$0.0003} \\
 &w/o CCA  &0.0664$\pm$0.0004  &0.1028$\pm$0.0006  &0.0355$\pm$0.0004  &0.0449$\pm$0.0005  \\
 &w/o UIA  &0.0725$\pm$0.0003  &0.1107$\pm$0.0005  &0.0390$\pm$0.0005  &0.0491$\pm$0.0005  \\
 &w/o REG  &0.0687$\pm$0.0005  &0.1061$\pm$0.0006  &0.0365$\pm$0.0003  &0.0464$\pm$0.0004  \\
 &w/o Vision  &0.0744$\pm$0.0003  &0.1122$\pm$0.0007  & 0.0399$\pm$0.0005 &0.0496$\pm$0.0007  \\
 &w/o Text  &0.0675$\pm$0.0004  &0.1024$\pm$0.0005  &0.0369$\pm$0.0004  &0.0460$\pm$0.0004  \\ \bottomrule
\end{tabular}
\label{tab:ablation-seed}
\vspace{-3mm}
\end{table*}

\subsection*{Hyper-parameter Settings}
For the multimodal encoder ($MMEnc$), we adopt the BEiT3-base architecture as its backbone, with the same network hyper-parameters. In this \textit{pre-training} stage, we set the input image resolution to $224^2$, and set the input text format to ``title+brand+description'' which can be directly obtained from Amazon items' metadata. When pre-training $MMEnc$, we set the learning rate to 0.0005 with batch size 1024, and set the optimizer to AdamW.
When performing recommendation training, the detailed hyper-parameters are presented in Table~\ref{tab:appendix-hyper}, with some basic settings derived from the MMRec. The values in $\{\ldots\}$ represent that we conduct the grid search on these values to find the best hyper-parameter.

\subsection*{Analysis on Parameter Amount and Inference Speed}
Due to the huge number of items and users, the structure's complexity in multimodal recommendation can significantly influence its feasibility to deployment. Therefore, we present the parameter amount and inference speed of AlignRec and three well-performed baselines in Table~\ref{tab:infer-time}. The inference time in this table means the total inference time over the test set. We can observe that the parameter amount of AlignRec is smaller than current baselines, with better performance. This is because our framework mainly focuses on the alignment design, training and evaluation, with small and simple network structures. Our multimodal representations are also smaller and more effective than before.

\begin{table}[t]
    \centering
    \caption{Results on parameter amount and inference speed on four methods.}
    \vspace{-3mm}
\begin{tabular}{c|c|cc}
\toprule
Dataset & Method & \# Parameter & Inference Time \\ \hline
\multirow{3}{*}{Baby} &AlignRec  &\textbf{7.17M} &\textbf{2.15s}   \\
 &BM3  &31.57M  &2.32s \\
 &MGCN  &35.58M  &2.40s  \\ 
 &FREEDOM  &33.56M  &2.22s  \\ \hline
\multirow{3}{*}{Sports} &AlignRec  &\textbf{17.60M} &\textbf{4.04s}   \\
 &BM3  &82.98M  &4.22s \\
 &MGCN  &87.00M  &4.26s  \\ 
 &MGCN  &85.96M  &4.13s  \\ \bottomrule
\end{tabular}
\label{tab:infer-time}
\end{table}

\subsection*{Mean and STD Values in Ablation Study}
To further validate the effectiveness of our framework, we conduct experiments on the ablation study with 5 seeds, and show the mean and std values in Table~\ref{tab:ablation-seed}. We can observe that each component contributes to AlignRec.

\subsection*{Further Hyper-parameter Analysis}

\begin{figure}[t]
    \centering
    \includegraphics[width=0.7\linewidth]{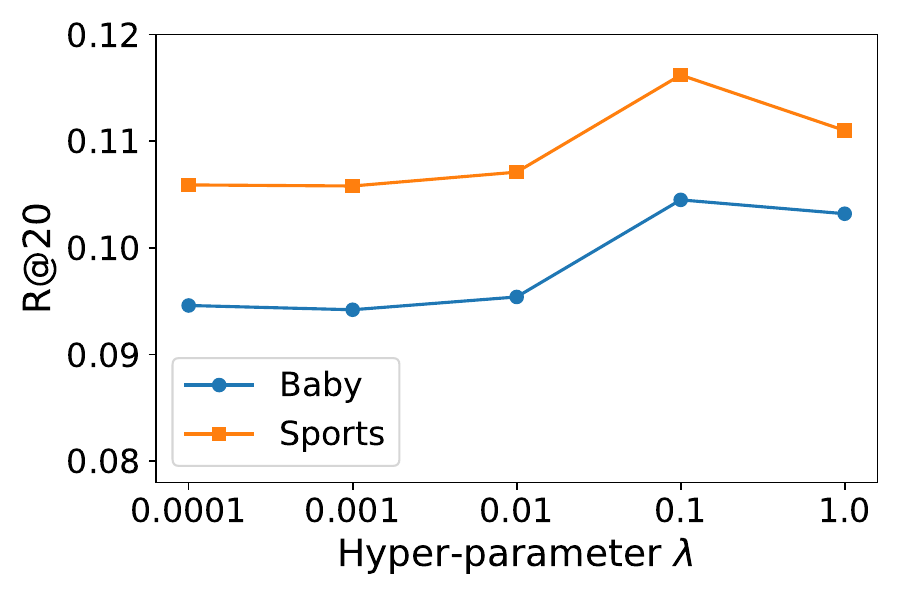}
    \caption{Hyper-parameter study on the weight $\lambda$.}
    \label{fig:hyper-lambda}
\end{figure}

\paragraph{Importance weight $\lambda$} It controls the importance of the similarity-based regularizer, on Amazon Baby and Sports dataset. As illustrated in Fig.~\ref{fig:hyper-lambda}, we find that when this weight is very small, it cannot effectively control the similarity among content modalities, so the content modality representation will be optimized towards the ID-based representation without considering the content similarity, losing its own knowledge. Therefore, the performance is relatively bad. On the contrary, if $\lambda$ is large (\eg, $\lambda=1$), the whole optimization process will mainly focus on this task, ignoring other alignment tasks. As a result, the performance also decreases. In general, $\lambda$ reaches the best performance around 0.1.

\paragraph{Number of layers $l$ in multimodal encoder} It determines the number of transformer layers in the multimodal encoder module, and is related to the scaling law of the multimodal recommendation. We change $l$ in $\{12,16,20\}$ ($l=12$ is the default setting in AlignRec), and the results are presented in Table~\ref{tab:scaling-law}. We can observe that when $l$ becomes larger, the performance slightly decreases. A larger $l$ represents more complicated networks in our multimodal encoder, which requires more data and more training time.

\begin{table}[t]
    \centering
    \caption{Hyper-parameter study on the number of layers $l$ in the multimodal encoder.}
    \vspace{-3mm}
\begin{tabular}{c|c|cccc}
\toprule
Dataset & $l$ & R@10 & R@20 & N@10 & N@20 \\ \hline
\multirow{3}{*}{Baby} &12  &0.0674  &0.1046 &0.0363 &0.0458  \\
 &16  &0.0679  &0.1052  &0.0370  &0.0463  \\
 &20  &0.0652  &0.1027  &0.0349  &0.0445  \\ \hline
\multirow{3}{*}{Sports} &12  &0.0758  &0.1160  &0.0414 &0.0517 \\
 &16  &0.0750  &0.1154  & 0.0409 &0.0512  \\
 &20  &0.0732  &0.1129  &0.0401  &0.0503  \\ \bottomrule
\end{tabular}
\label{tab:scaling-law}
\end{table}

\end{document}